\documentclass[preprint]{aastex}
\usepackage{graphicx,amsmath}

\newcommand\lta{\mathrel{\hbox{\raise 0.6 ex \hbox{$<$}\kern
                   -1.8 ex\lower .5 ex\hbox{$\sim$}}}}
\newcommand\gta{\mathrel{\hbox{\raise 0.6 ex \hbox{$>$}\kern
                   -1.7 ex\lower .5 ex\hbox{$\sim$}}}}
 
\newcommand{\scrbox}[1]{\ensuremath{{\mbox{\scriptsize #1}}}}

\newcommand{\teff}{{\ensuremath{T_{\scrbox{eff}}}}}
\newcommand{\Msol}{\ensuremath{\,\mbox{\it M}_{\odot}}}
\newcommand{\Mstar}{\ensuremath{\it \,M_{*}}}

\newcommand{\MS}{main--sequence}
\newcommand{\gr}{\ensuremath{g_{\scrbox{rad}}}}

\newcommand{\DM}{\ensuremath{ \log \Delta M/M_{*}}}

\newcommand{\K}{\mbox{K}}

\newcommand{\mix}{_{\rm{mix}}}

\newcommand{\ad}{_{\rm{ad}}}
\newcommand{\rad}{_{\rm{rad}}}

\shortauthors{Michaud et al.}
\shorttitle{Diffusive Stellar Models for Solar Abundances}

\begin{document}

\title{Models for Solar Abundance Stars with Gravitational Settling
and Radiative Accelerations: Application to M$\,$67 and NGC$\,$188}

\author{G.~Michaud, O.~Richard\altaffilmark{1}, \& J.~Richer}
\affil{D\'epartement de Physique, Universit\'e de Montr\'eal,
       Montr\'eal, PQ, H3C~3J7}
\email{michaudg@astro.umontreal.ca, Olivier.Richard@graal.univ-montp2.fr,
       jacques.richer@umontreal.ca}
 
\and

\author{Don A.~VandenBerg} 
\affil{Department of Physics \& Astronomy, University of Victoria, 
       P.O.~Box 3055, Victoria, BC, V8W~3P6, Canada}
\email{davb@uvvm.uvic.ca}

\altaffiltext{1}{GRAAL UMR5024, Universit\'e Montpellier II,
                 CC072, Place E. Bataillon,
                 34095 Montpellier Cedex,
                 France}

\begin{abstract}
Evolutionary models taking into account radiative accelerations,
thermal diffusion, and gravitational settling for 28 elements, 
including all those contributing to OPAL stellar opacities, have been
calculated for solar metallicity stars of 0.5 to 1.4 \Msol{}.  The Sun
has been used to calibrate the models.  Isochrones are fitted to the
observed color-magnitude diagrams (CMDs) of M$\,$67 and NGC$\,$188,
and ages of 3.7 and 6.4 Gyr are respectively determined.  Convective
core overshooting is {\it not} required to match the turnoff morphology
of either cluster, including the luminosity of the gap in M$\,$67,
because central convective cores are larger when diffusive processes
are treated.  This is due mainly to the enhanced helium and metal
abundances in the central regions of such models.  The observation of
solar metallicity open clusters with ages in the range 4.8--5.7 Gyr
would further test the calculations of atomic diffusion in central
stellar regions: according to non-diffusive isochrones, clusters should
not have gaps near their main-sequence turnoffs if they are older than
$\approx 4.8$ Gyr, whereas diffusive isochrones predict that gaps
should persist up to ages of $\approx 5.7$ Gyr.

Surface abundance isochrones are also calculated.  In the case of M$\,$67
and NGC$\,$188, surface abundance variations are expected to be small.
Abundance differences between stars of very similar \teff{} are
expected close to the turnoff, especially for elements between P and Ca.
Moreover, in comparison with the results obtained for giants, small
generalized underabundances are expected in \MS{} stars.  The lithium to beryllium 
ratio is discussed briefly and compared to observations.  The inclusion
of a turbulent transport parametrization that reduces surface abundance
variations does not significantly modify computed isochrones.  
\end{abstract} 


\today

\section{ASTROPHYSICAL CONTEXT}
\label{sec:intro}
The open clusters M$\,$67 and NGC$\,$188 have about the solar metallicity,
bracket the solar age, and have turnoff stars only a few hundred degrees
hotter than the Sun.  As such, they are interesting testing grounds for the
effects of atomic diffusion on age determinations and surface abundances,
since, in the case of the Sun, there is now ample evidence from heliosismology
that atomic diffusion has reduced the surface He abundance (\citealt{GuzikCo92};
\citealt{GuzikCo93}; \citealt{ChristensenDalsgaardPrTh93}; \citealt{Proffitt94};
\citealt{BahcallPiWa95}; \citealt{GuentherKiDe96}; \citealt{RichardVaChetal96};
\citealt{BrunTuZa99}).  Indeed, diffusive processes presumably also cause small
underabundances of metals in the Sun: these are caused mainly by gravitational
settling, but are also modified by radiative accelerations (\gr{}), which are
predicted to be especially important at the end of the \MS{} phases of
solar-type stars \citep{TurcotteRiMietal98}.  What abundance anomalies are
then to be expected in the turnoff stars of M$\,$67, which are $\sim 400$ K
hotter than the Sun (\citealt{HobbsTh91}), and in those of NGC$\,$188, which are
$\sim 100$ K hotter (\citealt{HobbsThRo90})?  As the cluster turnoff stars
are expected to have smaller surface convection zones, they may show larger
effects of atomic diffusion than the Sun.

On the other hand, since the radius and age of the Sun are used to calibrate
the mixing length and assumed initial He abundance, this normalization may
eliminate the effects of diffusion on age determinations.  Whereas an
$\approx 10$\% reduction in age at a given turnoff luminosity --- compared
with the predictions of models that neglect diffusion --- was derived by
\citet{VandenBergRiMietal2002} from the diffusive models for Population II
stars computed by \citet{RichardMiRietal2002}, it is not clear that
a similar reduction should be expected in the case of Pop.~I stars.  In
addition, there may be some important differences in the morphologies of the
diffusive and non-diffusive isochrones in the age range where a transition
is made between isochrones that have a gap near the main-sequence turnoff and
those which do not.  One naively expects that both the sizes of convective
cores in models for main-sequence stars of a given mass, and the predicted
mass marking the transition between stars that have convective and radiative
cores on the main sequence, will depend (to some extent) on whether or not
diffusive processes are treated.  (For instance, the concomitant increase
in opacity with the settling of Fe in the cores of stars would tend to
enhance convective instability.)

In this regard, we note that the first studies of NGC$\,$188
(\citealt{Sandage62}; \citealt{EggenSa69}) concluded that it has a gap
near the top of its main-sequence on the $(V,\;B-V)$-diagram reminiscent
of that seen in M$\,$67.  \citet{McClureTw77} carried out a statistical
test of the photographic photometry that they obtained for the same cluster,
and confirmed the existence of the gap in the color-magnitude diagram (CMD)
that was constructed for stars within Ring I on Sandage's original finder
chart.  Curiously, no statistically significant evidence for a gap was
found if their CMD included stars in Sandage's Ring II; but McClure
\& Twarog concluded that contamination by field stars was almost certainly
much more severe in the outer ring and that, if it were possible to
remove the field stars, ``the gap would be obvious".  The
 proper-motion membership study of \citet{DinescuGiAletal96}
does not shed any light on this problem (because of the very large scatter
in their CMD fainter than $V=15$: the gap is located at $V\approx 15.5$
according to McClure \& Twarog).  However \citet{PlataisKoMaetal2003}
provide a well defined CMD down to $V= 20$, with no indication of a turnoff gap.

Because there is no obvious indication of a gap in subsequent CMDs for
NGC$\,$188 (e.g., \citealt{Kaluzny90}; \citealt{CaputoChCaetal90};
\citealt{SarajediniHiKoetal99}), and because the best-fitting isochrones for
current best estimates of the cluster distance and reddening do not
predict a gap at the turnoff $M_V$ (see the aforementioned papers),
its existence is considered by many to be quite doubtful.
However, the models that have been compared with the cluster CMD have
not taken gravitational settling and radiative accelerations into account.
If it were shown that diffusive isochrones do, in fact, predict a
main-sequence gap, not necessarily for NGC$\,$188 but for any range in
age where a gap is not predicted by models that neglect diffusion, this
would be an important development that would motivate a search for open
clusters within the requisite age range to further test our understanding
of stellar physics.

There are many other questions that need to be addressed.  In particular,
does the same turbulence parametrization that Richard et al.~(2002) used,
in conjuction with diffusion physics, to explain the Li abundances in
field halo stars, also lead to good agreement between the predicted and
observed Li abundances in solar-type stars, as well as in those that were
in the Li gap at the age of the Hyades \citep{Balachandran95}?  The
surface abundances of Li in M$\,$67 solar-type stars (see Fig.~7 of
\citealt{MartinBaPaetal2002}) vary from star to star at a given \teff{},
which suggests that mixing processes below the surface convection zone
vary from star to star at a given mass.  Did the turbulence differ only
in the early stellar histories of the cluster stars or is it still
different between one solar twin and another?  To what extent is this
confirmed by abundance anomalies of other species and do such anomalies
affect theoretical isochrones?  Furthermore, what abundance anomalies of
Fe, Li, C, and O could be caused by diffusion and are they observed
\citep{BarrettBoKietal2001}?

In NGC$\,$188, the surface Li abundance in solar-type stars appears to be
consistent with a single-valued function of $\teff$ just as in the Hyades
(see \citealt{RandichSePa2003}).  Furthermore, even though NGC$\,$188 is
older than M$\,$67 by $\sim 2$--$3\times 10^9$ years (e.g., Sarajedini et
al.~1999), the Li abundance in its G-type stars is comparable with the
largest abundances measured in M$\,$67 stars having similar 
colors/temperatures.  Could a simpler model account for the Li 
observations in the Hyades and NGC\,188 than in M\,67? Pre--\MS{} evolution
could be largely responsible for the Li destruction in the Hyades and
NGC\,188 (see, for instance, \citealt{ProffittMi89, PiauTu2002}). Finally, we note
that the relative abundances of Li and Be along the subgiant branch of M$\,$67
has been evaluated in models including gravitational settling by \cite{SillsDe2000}.  However, what are the ratios of
the Li and Be abundances to be expected from diffusion if \gr{} are also taken
into account? 

In this paper, after a very brief description of the calculations in \S \ref{sec:calcul},
the chemical composition expected on the surfaces of stars of M\,67 and NGC\,188 will be discussed in \S \ref{sec:composition} and Li/Be ratios in \S \ref{sec:li_be}.
The effect of atomic diffusion on central convective cores is analyzed in detail in \S \ref{sec:CZ}.
The effect of diffusion on isochrones is discussed in  \S \ref{sec:iso} and   these results are applied to M\,67 and NGC\,188 in \S \ref{sec:opencl}.
The main conclusions are summarized in \S \ref{sec:conclusion}.  Throughout this paper the emphasis is on calculations in the presence of atomic diffusion 
and, in some cases, of turbulent transport with the same parametrization as used for Pop II stars by \citet{RichardMiRietal2002}.
The discussion of potential star--to--star variations of turbulent transport to explain Li abundance spread at a given \teff{}
 is left to a paper in preparation.

\section{CALCULATIONS}
\label{sec:calcul}

The models were calculated as described by \citet{TurcotteRiMietal98}
and \citet{RichardMiRi2001}.  They were assumed to be chemically
homogeneous on the pre-main sequence with a solar abundance mix, and
relative concentrations as defined in Table 1 of \citet{TurcotteRiMietal98}. 
The radiative accelerations are from \citet{RicherMiRoetal98} with
the correction for redistribution from \citet{GonzalezLeAretal95} and
\citet{LeblancMiRi2000}. The atomic diffusion coefficients were taken
from \citet{PaquettePeFoetal86} (see also \citealt{MichaudPr93}).
In all cases, the Krishna Swamy $T$--$\tau$ relation \citep{Krishna_Swamy66}
was used to derive the outer boundary condition for the pressure that is
needed to construct stellar models.  Semiconvection was included as
described in \citet{RichardMiRi2001}, following \citet{Kato66},
\citet{LangerElFr85}, and \citet{Maeder97}.

In \citet{TurcotteRiMietal98}, the solar luminosity and radius at the
solar age were used to determine the value of $\alpha{}$, the ratio of the
mixing length to the pressure scale-height in the usual mixing-length
theory (MLT) of convection, and of $Y_0$, the He concentration in the
zero-age Sun. The value of $Y_0$ mainly affects the luminosity while
$\alpha{}$ primarily determines the radius, through the depth of the
surface convection zone. The required value of $\alpha{}$ was found to
be slightly larger in the diffusive, than in the non-diffusive, models
because an increased value of $\alpha{}$ is needed to compensate for the
settling of He and the metals from the surface convection zone.  The
increase in $\alpha{}$ in the diffusion models of the Sun is thus
determined by the settling that occurs immediately below the solar
surface convection zone.

The value of $\alpha{}$ and the initial values of $Y_0$ and $Z_0$ that were
adopted in each of the three series of models computed for this study are given
in Table \ref{tab:parameters}, together with the accuracy with which they
represent the solar properties at the solar age.  We did not force
convergence as precisely as in \citet{TurcotteRiMietal98}, but the
convergence should suffice for the purposes of this paper.  The model
with atomic diffusion only is calculated using the same values of $Y_0$,
$Z_0$, and $\alpha{}$ as in \citet{TurcotteRiMietal98}, even though small
changes were since made to the code, such as a better treatment of some
interaction terms in the diffusion equations for the various species.
(This is what has caused a slight degradation of the convergence criteria.)
For the non-diffusive case, the values of $Y_0$, $Z_0$, and $\alpha{}$
tabulated by \citet{TurcotteRiMietal98} (Model B of their Table 2) were
calculated using tables of mean opacities while, here, the monochromatic
opacities were used even for models without diffusion.  This causes small
differences in the central regions of the solar model where CNO abundance
variations modify the opacity --- leading to slight differences in our
value of $\alpha{}$ (see Table \ref{tab:parameters}) compared with that
used in model B of \citet{TurcotteRiMietal98}.   

One series of models was calculated with the same turbulent transport
parametrization, labeled T6.09, that was found to minimize the surface Li
abundance changes in Pop.~II field stars (see
\citealt{RichardMiRietal2002} for both the definition of this turbulent
transport parametrization and its justification).  As may be seen from
Fig.~6 of \citet{RichardMiRietal2002}, that turbulent transport coefficient
approximately equals the He atomic diffusion coefficient at $\log T = 6.3$ 
and diminishes rapidly as $T$ increases further.
Because this is the temperature close to the bottom of the solar surface
convection zone, this level of turbulent transport does not affect solar
models significantly.  We have, in fact, verified that the same values of
$Y_0$, $Z_0$, and $\alpha{}$ are obtained for the calibrated solar
models that allow for only atomic diffusion, on the one hand, and atomic
diffusion plus T6.09 turbulence, on the other.  Since, furthermore, it is
the effect of adding turbulence to models with atomic diffusion that we
wish to study, the models with turbulence must have the same $Y_0$ and
$\alpha{}$ as those with atomic diffusion only.

\section{EVOLUTIONARY MODELS}
\label{sec:EVOLUTION}

In Figure~\ref{fig:hist} are shown, for a few of the calculated models,
the time dependence of \teff{} as well as of the depth of the surface
and central convection zones.  The data were taken from some of the
models without diffusion (top row) and from some with atomic diffusion
(bottom row).  The surface convection zone mixes to the surface the
abundances that are modified by atomic diffusion below the fully mixed
outer layers. The time dependence of the depth of the surface convection
zone determines the time dependence of the depth of the region where element
separation occurs.  In these models, the smallest convection zones (in
terms of the amount of mass that they contain) occur early in the evolution
and for a brief period just past the turnoff.  This is different from the
stars in low-metallicity Pop.~II globular clusters in which the mass in the
surface convection zone decreases throughout the \MS{} phase (see Fig.~1
of \citealt{RichardMiRietal2002}).

\subsection{Chemical composition}
\label{sec:composition}
Figure~\ref{fig:abundanc_iso} illustrates the variation in the surface
abundances of several species as a function of \teff{} at 3.7 Gyr, which is our
estimate of the age of M$\,$67 (see \S \ref{sec:opencl} below).  These ``surface
abundance isochrones" were calculated for an additional 19 species, but
only representative ones are shown: the other loci bear considerable
similarity to those that have been plotted.  The corresponding \gr{}
are illustrated in Figure~\ref{fig:gr} for B, Mg, P, Ti, Fe, and Ni.
(The \gr{} of He, Li, and Be are negligible in 1.3
\Msol{} models.)

The surface abundances of $^3$He and LiBeB are affected by both
diffusion processes and nuclear reactions.  The effect of nuclear
reactions on the surface abundances of these elements becomes evident
during the evolution of a star on the subgiant branch when dredge-up  
occurs.  Overabundances of $^3$He are predicted to appear as the star reaches $\teff\approx
5400$ K (the temperature where the surface abundance isochrone becomes
nearly vertical in Fig.~\ref{fig:abundanc_iso}).  At this point in the
star's evolution, the bottom of the surface convection zone reaches down to
regions where $^3$He has a concentration maximum produced during the \MS{}
stage \citep{Iben65}.  For LiBeB, underabundances are expected at $\teff\lta
5800$~K for Li and Be and $\lta 5600$ K for B (see
Fig.~\ref{fig:abundanc_iso}): this occurs as the bottom of the surface
convection zone reaches the regions where Li, Be, and B burn.  The other
abundance variations are caused by atomic diffusion.

\begin{figure}[p]
\centerline{
\includegraphics[width=\textwidth]{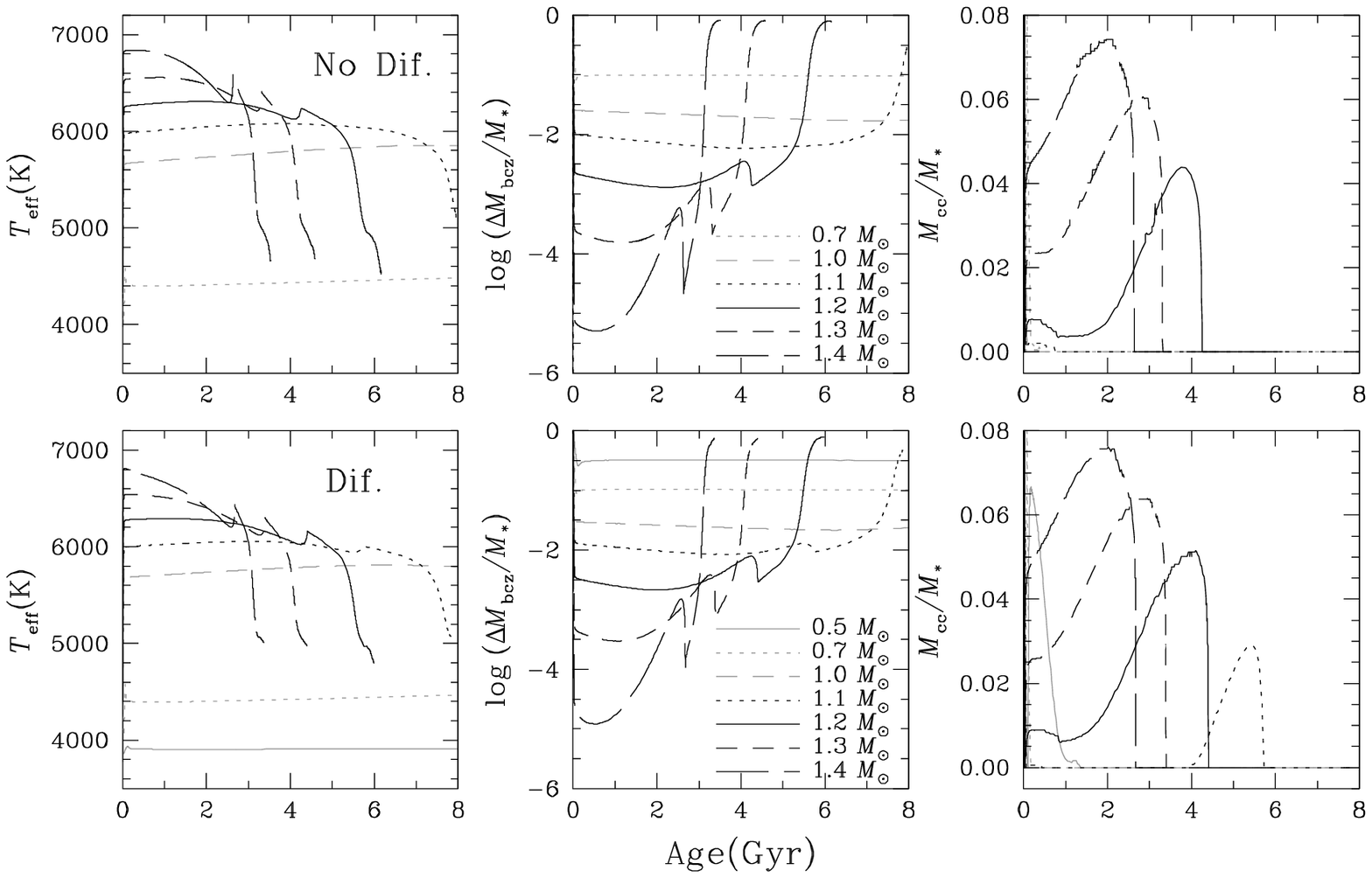}}
\caption{Properties of a few of the calculated models without diffusion
(top row) and with atomic diffusion (bottom row) as a function of time.
Surface abundances are mainly determined by diffusion processes occurring
immediately below the surface convection zone (center panels).  The mass
of the central convective core is shown in the right-hand panels.  
While the 1.1 \Msol{} model with diffusion has a central convective core, that
without diffusion does not.
}
\label{fig:hist}
\end{figure}

\begin{figure}[p]
\centerline{
\includegraphics[width=\textwidth]{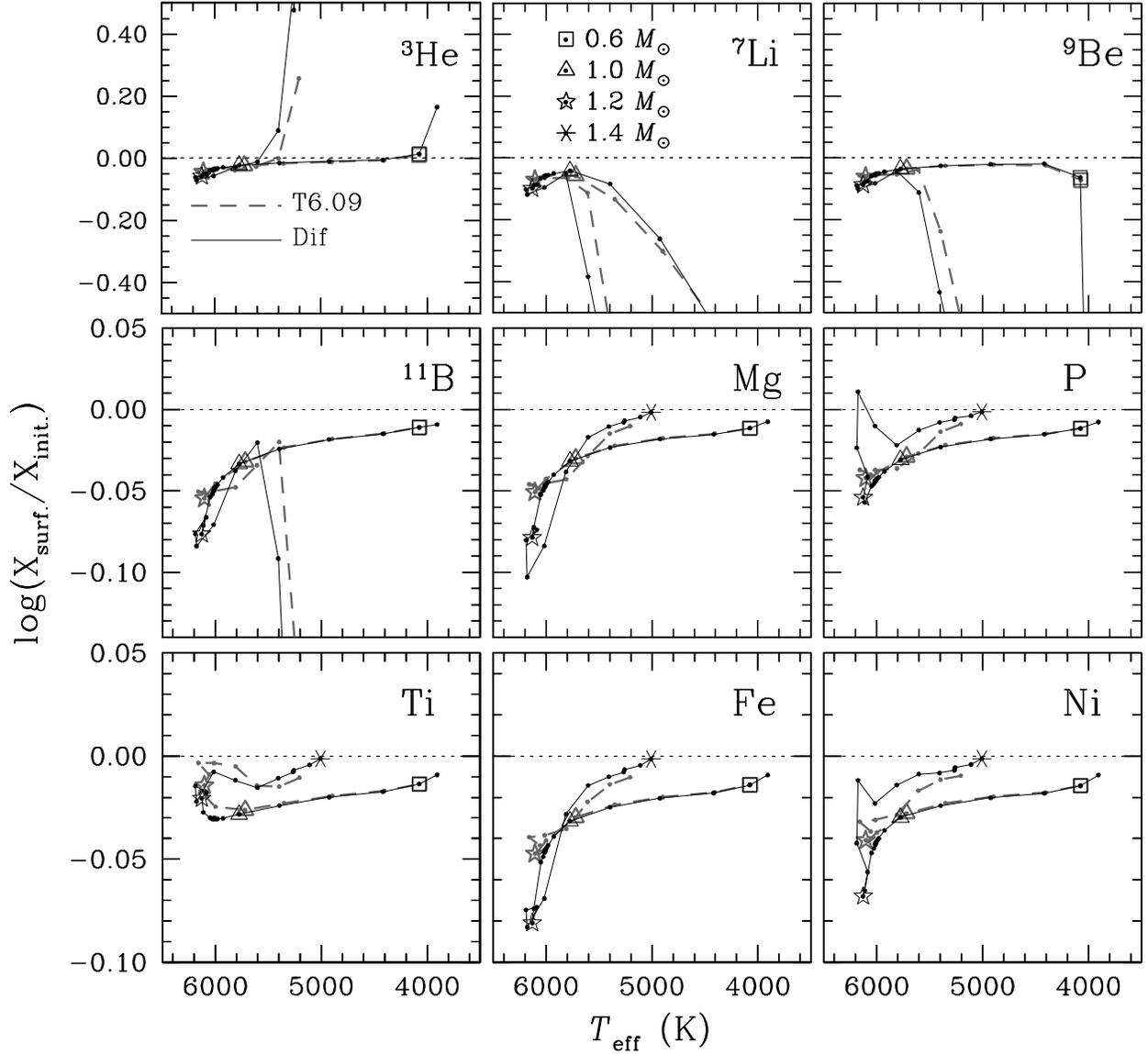}}
\caption{Chemical abundances at the surface of solar metallicity stars
at 3.7 Gyr in models with atomic diffusion and in those with
T6.09 turbulence.  Such surface abundance isochrones were calculated
for 28 species, but only representative ones are shown.  The smallest
mass star is always at the extreme right of each curve. Special characters are used for a few stellar masses 
and they are identified on the figure.
}
\label{fig:abundanc_iso}
\end{figure}

\begin{figure}
\centerline{
\includegraphics[width=\textwidth]{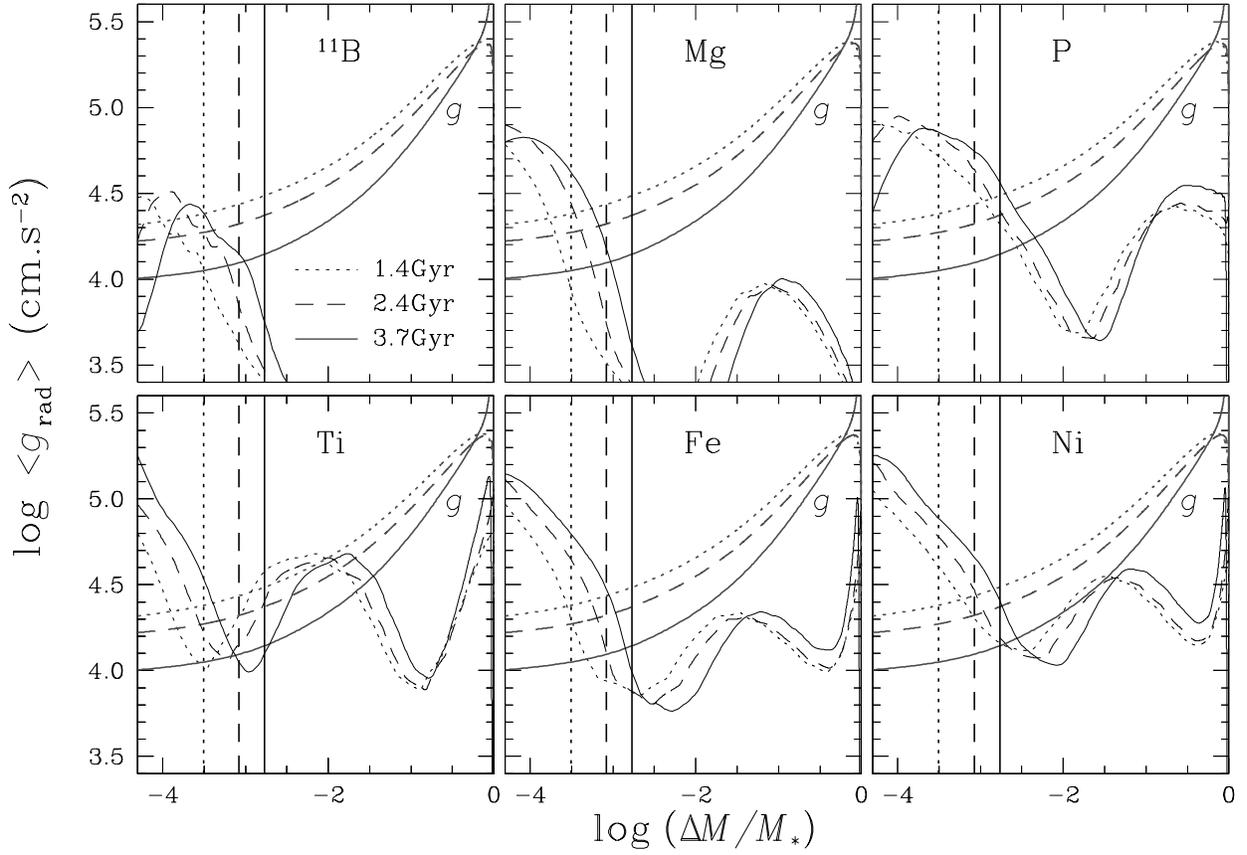}}
\caption{Radiative accelerations in a 1.3 \Msol{} solar metallicity star
at 1.4, 2.4, and 3.7 Gyr. The bottom of the surface convection zone at
each epoch is indicated by a vertical line of the same type. Gravity ($g$)
is plotted in each panel of the figure.  The \gr{} of Li and Be are not
shown because they are always smaller than that of B below the surface
convection zone and so do not have a significant impact on Li and Be
abundances.
}
\label{fig:gr}
\end{figure}

In the model with atomic diffusion only, the surface abundance
variations as a function of time are directly related to the depth 
of the surface convection zone (see Fig. \ref{fig:hist}): overabundances 
normally appear when $ \gr{} \ge g$ immediately below the surface
convection zone.  When the reverse is true, underabundances generally
appear at the surface.  The detailed results may be understood by
remembering that, for those elements whose $\gr$ is small, the surface
abundance decreases approximately as
\begin{equation}
\label{eq:e_t}
\exp(-t/\theta)
\end{equation}
 where
\begin{equation}
\label{eq:theta}
\theta \simeq 2.3\times10^{11}  (\Delta M/\Msol)^{0.545} yr
\end{equation}
(for helium, with similar expressions for other species; see
\citealt{Michaud77a}).  The precise value of the multiplying constant
varies slightly with stellar mass, but more so with the atomic weight
of each element and its charge\footnote{In this paper, $\Delta M$
always represents the mass of the spherical shell {\em outside\/} a
certain radius.  Furthermore, in the above equation, this mass is
assumed to be mixed (for instance, by convection).}. When $\theta$ is
smaller than the age of the star, the abundance reached is a very
sensitive function of the mixed mass. 

The \gr{} for B below the surface convection zone (see Fig.~\ref{fig:gr})
is always smaller than gravity by at least a factor of two.  Consequently,
it has no more than a small effect (if any) on the boron concentration, which
is mainly determined by gravitational settling until the bottom of the surface
convection zone reaches the temperature where B burns.

The Mg abundance has a \teff{} variation typical of species from O to
Si: it is caused by \gr{} (see Fig. \ref{fig:gr}) being much smaller
than gravity below the surface convection zone, throughout the evolution
of the stars in the relevant mass range.  Elements from O to Si have
\gr{} with a similar \DM{} dependence\footnote{To see the variation of
all \gr{} with $T$, reference may be made to Fig.~1 in the study by
\citet{RicherMiRoetal98}.}.  All of these elements settle by gravitation
below the surface convection zone and they have the largest underabundances
at the end of the \MS{} phase.  The underabundances are larger in the more
massive stars because they have smaller surface convection zones (see
Fig.~\ref{fig:hist}).

P and Ti represent all elements between P and Ti (except for S, which
is more like Mg).  Their \gr{} are slightly larger than gravity for a
significant mass interval below the convection zone (see Fig.~\ref{fig:gr}).
The mass interval where \gr{} is large varies from P to Ti.  As the atomic
number of the species increases, the larger values of \gr{} shift to a
greater depth.  Very small overabundances may appear at the turnoff, but at a
later epoch only in the more massive stars considered here.  For most species,
the effect of \gr{} is merely to reduce the expected underabundances in the
hotter stars.

Fe is representative of species of the Cr, Mn, Fe group.  For all of them,
\gr{} is continuously smaller than gravity below the surface convection
zone, but not by as large a factor as for Mg.  Consequently, the predicted
underabundances are not as large either.  Finally, Ni is supported below
the convection zone in the hotter stars considered. 
 
In the presence of T6.09 turbulence, one expects underabundances of the
metals at the $\sim 6$\% level in stars between 4000 and 5000 K, progressively
increasing to $\sim 12$\% in stars of 6000 K.  Only underabundances are
predicted because the T6.09 turbulence mixes deep enough in the star (down
to $\DM \simeq -2$) for \gr{} never to play a dominant role. The $\gr{}$
still limit the underabundances of a number of species and, in particular, of
Ti, as may be seen in Figure \ref{fig:abundanc_iso}.

\begin{figure}
\centerline{
\includegraphics[width=\textwidth]{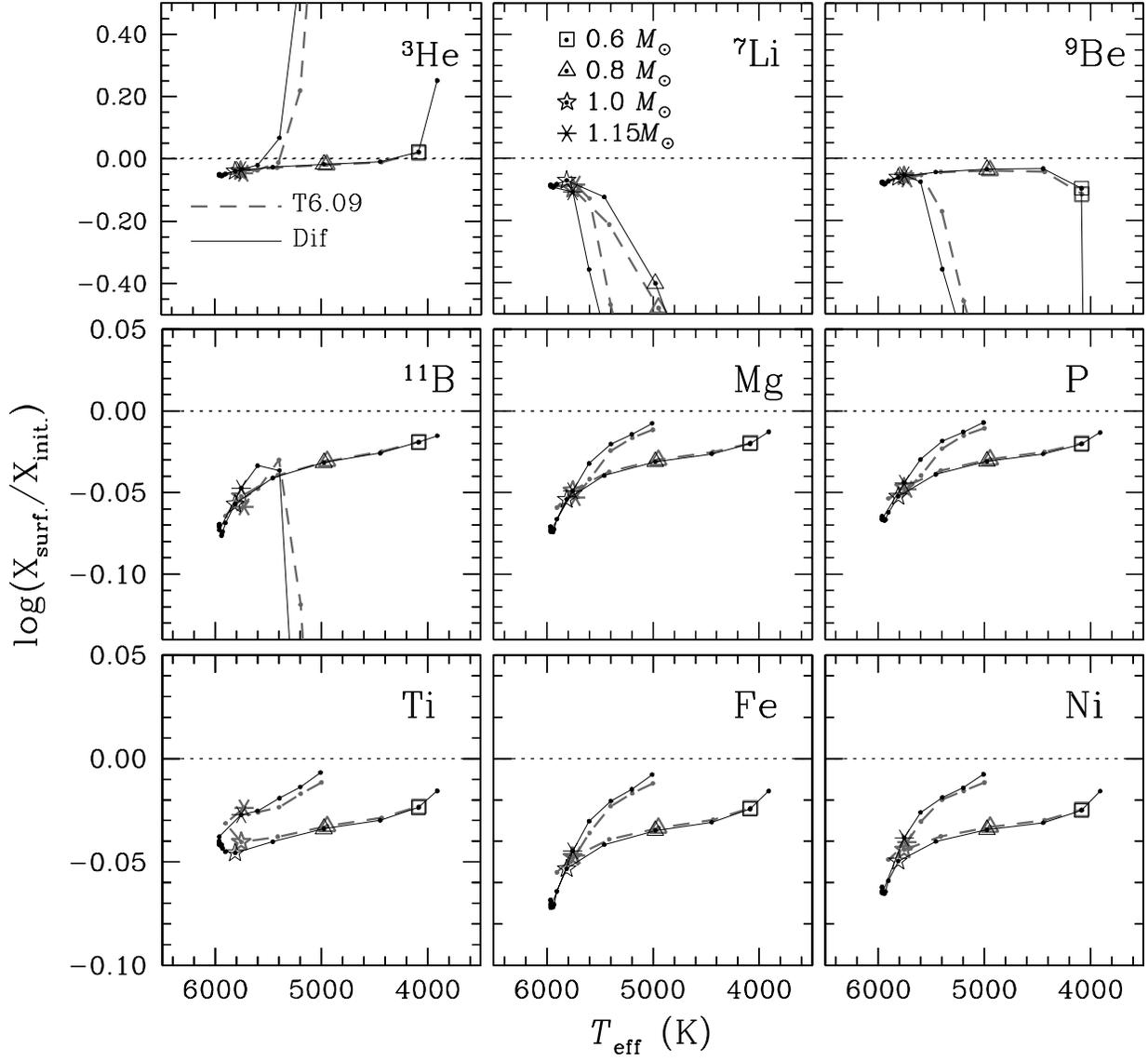}}
\caption{Same as in Fig. \ref{fig:abundanc_iso} but 
at 6.4 Gyr as appropriate for NGC\,188. 
}
\label{fig:abundanc_188}
\end{figure}

Similar results are shown in Figure \ref{fig:abundanc_188} at the epoch
corresponding to the age determined below for NGC\,188, 6.4 Gyr.  The
abundance anomalies are larger at a given \teff{} than those 2.7 Gyr
earlier because of the longer time available for gravitational settling
(Eq.~\ref{eq:e_t}).  However, the stars with the largest \teff{} at 3.7 Gyr
have, at 6.4 Gyr, evolved away from the main sequence.  They were the ones with the
largest spread of anomalies of P, Fe, and Ni, at a given \teff{}, in
Fig.~\ref{fig:abundanc_iso}.  The spread of abundance anomalies at the turnoff 
in Fig.~\ref{fig:abundanc_188} is much smaller due to the reduced importance
of \gr{}: only for Ti is the effect of \gr{} still clearly visible. 
Similarly, at that age, T6.09 turbulence has very little effect since surface
convection zones extend to $T \simeq 10^6$ \K{} or $\sim 10$ times deeper than
at 3.7 Gyr (see Fig. \ref{fig:hist}).

\subsection{The Li/Be ratio in M\,67}
\label{sec:li_be}

\begin{figure}
\centerline{
\includegraphics[width=0.7\textwidth]{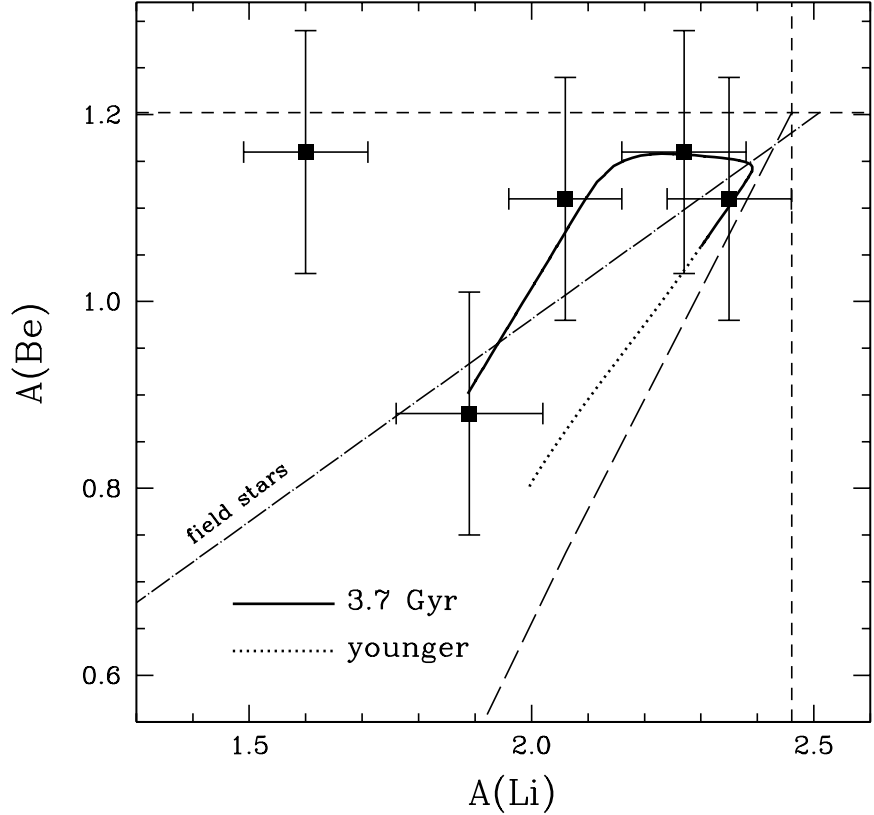}}
\caption{Isochrone of the lithium to beryllium abundance ratio in M\,67 stars with $\teff \ge 5500$\,K from models with atomic diffusion (solid curves
of Li and Be on Fig. \ref{fig:abundanc_iso}).
The left segment of the solid curve represents  stars starting on the subgiant branch while the right segment
indicates \MS{} stars and the nearly horizontal one denotes stars at turnoff.  The dotted part of the curve was occupied
at earlier times by stars that have now evolved to the subgiant or giant evolutionary state.
The horizontal and vertical dashed lines indicate the initial values used in the calculations. 
The data points are from \citet{RandichPrPaetal2002} and their quoted error bars are plotted.  
 For comparison purposes, 
the long dashed curve gives the evaluation of Li/Be due to atomic diffusion
 by \citet{SillsDe2000}  (adjusted to have the same zero age \MS{} values as we used), while the dot dashed 
curve is their fit to the observed Li/Be ratio in field stars.  
}
\label{fig:li_be_m67}
\end{figure}

 \citet{SillsDe2000} have calculated the Li/Be abundance ratio with a number of evolutionary models 
that included either turbulent transport, or atomic diffusion, or no transport process (dubbed \emph{standard models}) and
have presented a comparison of the different results in their Fig.\,5.  They used these results
to determine the relative importance of those  transport processes.
In their models with atomic diffusion only, they obtain approximately a factor of 100 reduction of the Be abundance 
at the same time as a factor 
of 30 reduction of the Li abundance or $X(\mbox{Be}) \sim [X(\mbox{Li})]^{4/3}$.  This is to be 
compared to the results shown on Figure \ref{fig:li_be_m67}
of this paper.  On the leg of the isochrone corresponding to \MS{} stars, one has 
$X(\mbox{Be})\sim [X(\mbox{Li})]^{4/5}$ while on the leg corresponding to subgiants, one has 
$X(\mbox{Be})\sim X(\mbox{Li})$.   Combining 
the two segments would give approximately $X(\mbox{Be})\sim [X(\mbox{Li})]^{1/2}$ but with a large dispersion 
 which is in agreement with
the Li/Be ratios observed in  field stars and discussed by  \citet{SillsDe2000}.  Our results for the diffusion 
models are very different from theirs probably because of our more complete description of atomic diffusion
processes.

Observations of  Li and Be in M\,67 were recently made by \citet{RandichPrPaetal2002} and their Li/Be ratios are also
 plotted on Figure \ref{fig:li_be_m67}.  Some of their stars have V magnitudes that correspond
to stars just before turnoff while others are slightly above it. Four of the 
five observed points are compatible with  the model that includes only atomic diffusion processes. 
Given the error bars, the agreement could be considered satisfactory, however the star with the smallest Be abundance (S988) has
a magnitude corresponding to pre-turnoff stars and so should not be on that segment of the curve.

One may also note that the original Li abundance used in these calculations (see Fig.\,\ref{fig:li_be_m67})  is smaller
than usually believed to be appropriate in young solar metallicity clusters.  This may however be affected by 
 pre--\MS{} burning
 (see, for instance, \citealt{ProffittMi89, PiauTu2002}) which was neglected in this paper.  Explaining the range of 
Li abundances observed in cluster and field stars requires a discussion of processes competing with atomic diffusion. 
This is outside the scope of the present paper but will be part of a paper in preparation.

\subsection{Central convective cores and semiconvection}
\label{sec:CZ}
\begin{figure}
\centerline{
\includegraphics[width=0.8\textwidth]{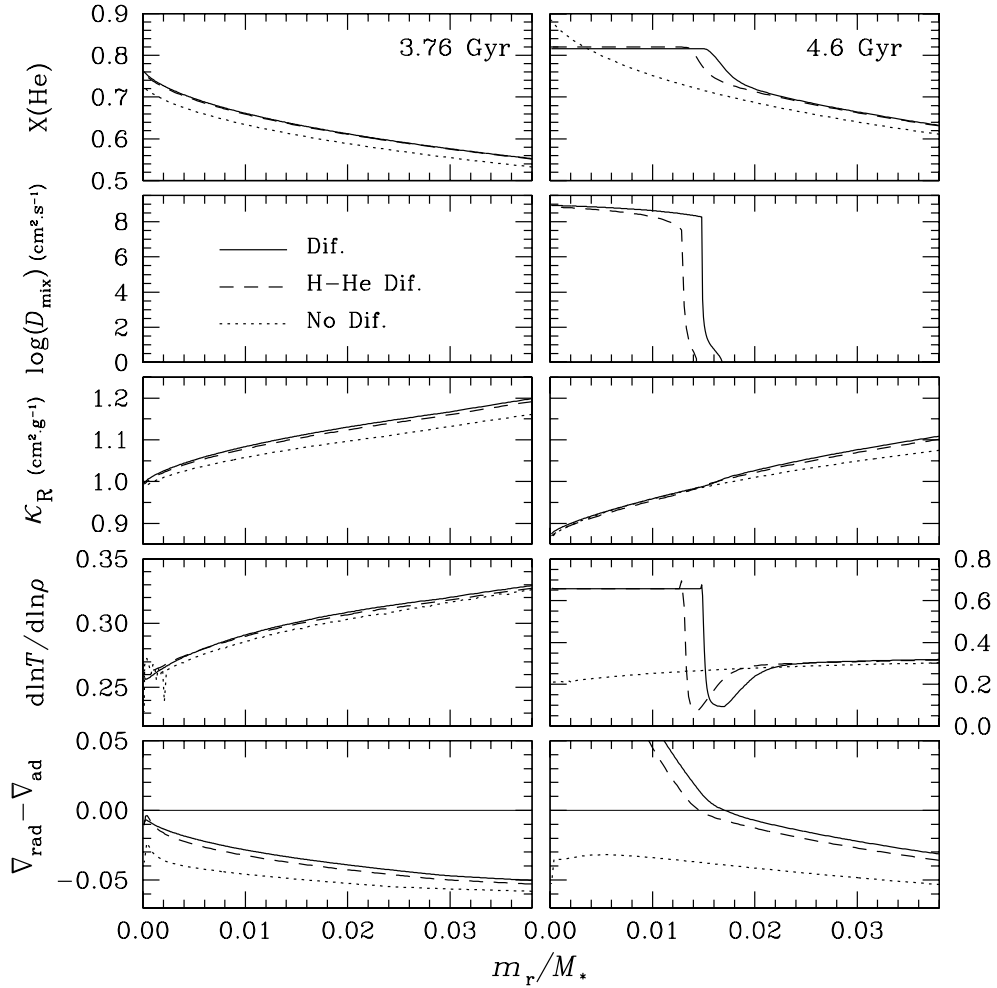}}
\caption{Important properties of three 1.10 \Msol{} models at $\simeq
3.76$ Gyr (left panels), just before the appearance of the central
convective cores in the models with diffusion. The {\it solid line} is the
model with atomic diffusion of all species, the {\it dashed line} is the model
with H-He diffusion but no diffusion of metals, while the {\it dotted line}
is the model without diffusion.  In the panels to the right are shown
the same properties when the convective cores are well developed in
the diffusive models (at 4.6 Gyr).  The same scales are used for the 
right- and left-hand panels, except for the $d\ln T/ d\ln \rho$ panel.
The mixing coefficient, $D\mix$, is caused by convection over the $m_r/\Mstar$
interval where $D\mix$ is nearly horizontal and equal to $10^8$ cm$^2$s$^{-1}$.
At the upper boundary of the convective core, it drops to $10^2$
cm$^2$s$^{-1}$ and then continues decreasing in the semiconvection zone.
While $X$(He) is constant in the convective core, there remains a
substantial He abundance variation in the semiconvective region.
From the plots in the bottom row, one sees that the radiative and
adiabatic gradients become equal approximately at the upper boundary
of the semiconvective zone, so that using the Schwarzschild, instead of
the Ledoux, criterion would actually lead to a larger convective core
than the Ledoux criterion used here.  It would include both the
convective core and its semiconvective extension.
}
\label{fig:kcentre1}
\end{figure}

One of the consequences of the use of diffusive models is to modify
the size of the central convective core (see Fig. \ref{fig:hist}), which
will be seen in \S \ref{sec:iso} to have a significant impact on the shape
of temperature--luminosity isochrones.  The convective core is larger
in the diffusive models of a given mass than in those that neglect
diffusion: 10\% larger at 1.3 \Msol{}, 20\% at 1.2 \Msol{}.  Moreover,
while the lowest mass \emph{non-diffusive} model with a convective core
is that for 1.14 \Msol{}, the lowest mass \emph{diffusive} model with
a convective core has a mass of 1.097 \Msol{}. This difference may be
understood by studying the central properties of 1.1 \Msol{} models.
Three different models are compared in Figure~\ref{fig:kcentre1}; (i)
our standard model with diffusion, (ii) that without diffusion, and
(iii) one with the diffusion of He but without the diffusion of
metals\footnote{Note that, only for this discussion, do we consider
a model with the diffusion of He but not of the metals.}.

The difference between the models with, and without, diffusion
originates from metallicity and He abundance variations.  Because of
the normalization to the current properties of the Sun (see
\citealt{TurcotteRiMietal98}), the initial $Y$ ($Y_0$) is 3\% larger
in the solar model with diffusion than in the one without diffusion.  
On the other hand, in order to have the observed value of the ratio
of surface metals to hydrogen ($Z_\odot/X_\odot$) at the solar age,
the initial value of $Z$ ($Z_0$) must be about 13\% larger in the 
original solar model with diffusion than in the one without diffusion
--- in order to compensate appropriately for the effects of atomic
diffusion during solar evolution (see Tables 2 and 6 of
\citealt{TurcotteRiMietal98}).  Consequently, our series of
non-diffusive models have smaller $Y_0$ and $Z_0$ than our series of
diffusive models (see Table \ref{tab:parameters}).  During the evolution,
the central values of $Y$ and $Z$ are further increased by 3\% to 4\% by
diffusion processes.  At an age of 3.76Gyr, the central values of $Y$
and $Z$ are consequently larger by about 7\% and 18\%, respectively,
in the diffusive compared to non-diffusive models of the same age. 

\begin{figure}
\centerline{
\includegraphics[width=\textwidth]{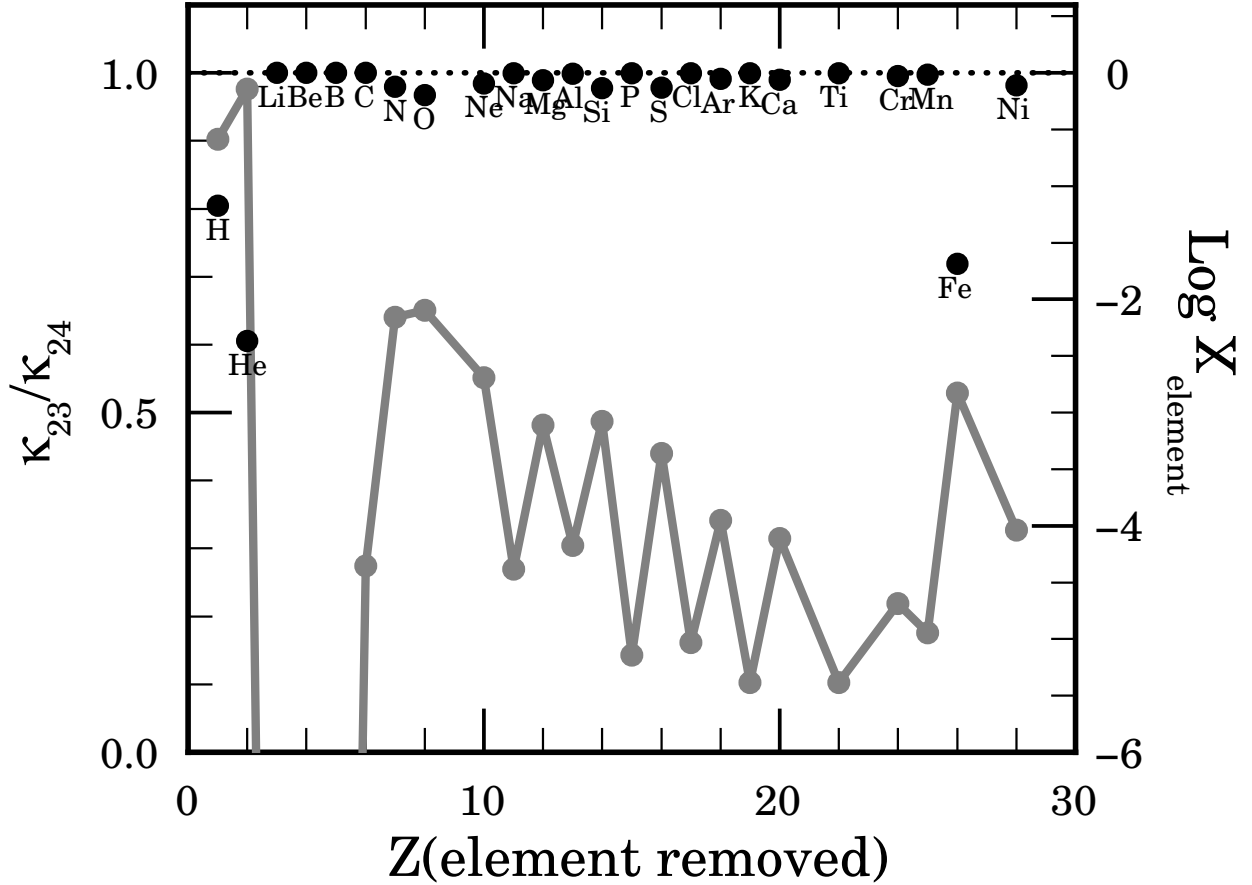}}
\caption{Evaluation of the contribution of each element to the 
Rosseland-averaged opacity in the 1.1 \Msol{} model with diffusion at
an age of 4.6 Gyr, just outside of the convective core (near $m_r/M_*
\simeq 0.02$, see Fig. \ref{fig:kcentre1}): the gray line and the
right-hand scale give the local mass-fraction abundance of each element.
Fe contributes to the opacity as much as H or He, and an increase in the
Fe abundance leads to a significant increase in the Rosseland opacity.
Since Fe contributes about one-third of the opacity, an 18\% increase
in the Fe abundance leads to a 6\% increase in the opacity.  For Fe,
which has not lost all of its electrons, the main contribution is from
bound-bound and bound-free transitions, while for H and He, the main
contribution is from free-free transitions.
}
\label{fig:kvsX}
\end{figure}

As may be seen from Figure \ref{fig:kvsX}, Fe contributes as much
to the Rosseland opacity as H or He, so that an 18\% increase in the
abundance of Fe leads to about a 6\% increase in Rosseland opacity at a given
$T$ and $\rho$.  Furthermore, a given mass of He contributes less to the
opacity than the same mass of H (because H and He contributions
to the opacity come mainly from their free electrons); with the result
that, as $Y$ increases, the opacity decreases.  The increase of He
abundance reduces from 6\% to  5\% the increase in opacity, at given $T$ and
$\rho$, caused by the 18\% increase of Fe abundance.  The effect may be
seen just before the appearance of convective cores in the left-hand
panel of Figure~\ref{fig:kcentre1}, where the opacity per gram is
approximately 4\% larger in the diffusive, than in the non-diffusive,
models (at $m_r/\Mstar =$ 0.038)\footnote{At a given $m_r/\Mstar$, the
$T$ and $\rho$ are not exactly the same in the three models because of
structural differences, which explains why the opacity increase is 4\%
in the models while it is 5\% at given $T$ and $\rho$.}.  After the
appearance of the convective core (right-hand panel), the opacity
outside the core is still larger in the diffusive, than in the
non-diffusive, models but structural changes wipe out the opacity
differences inside the core itself.

\begin{figure}
\centerline{
\includegraphics[width=\textwidth]{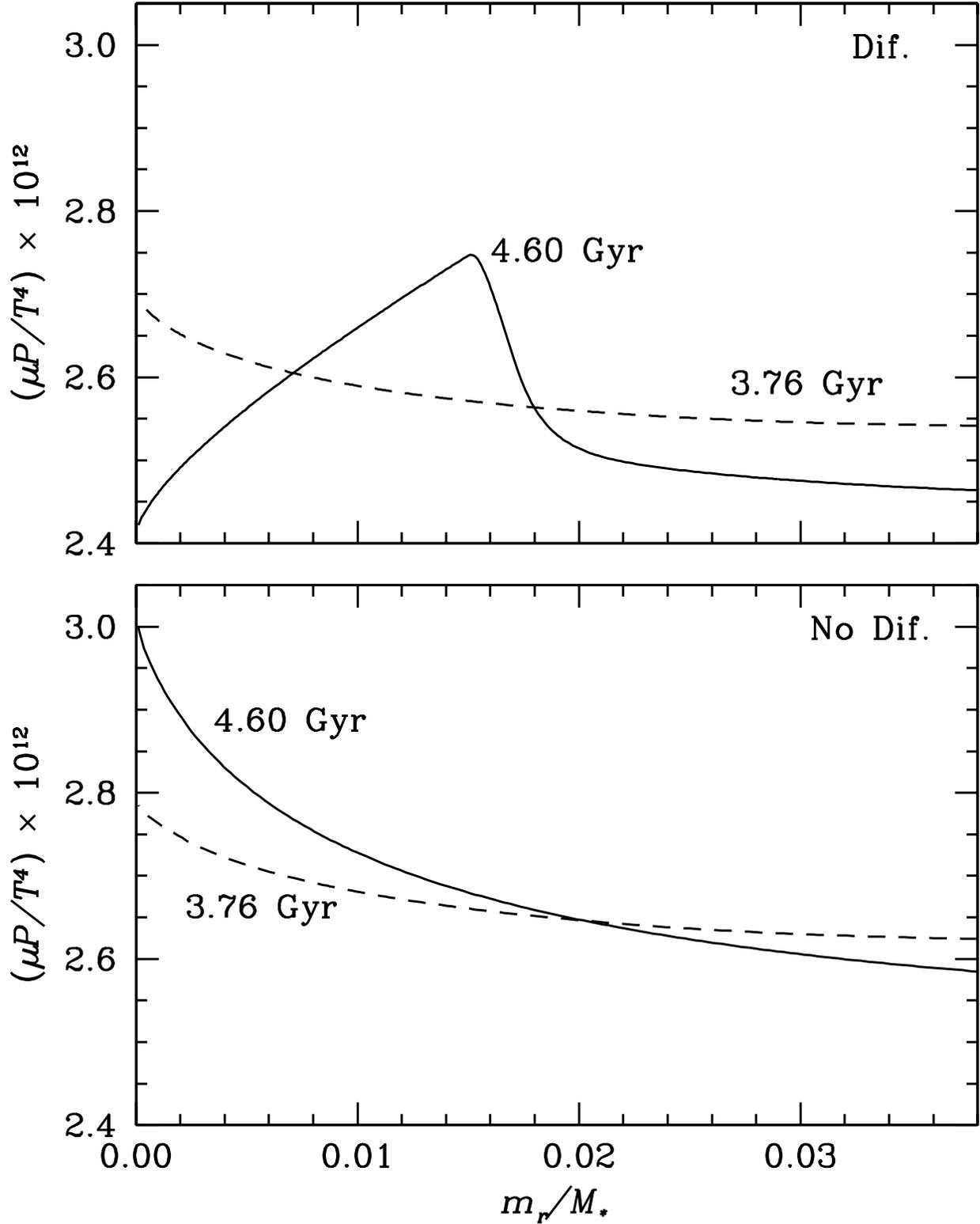}}
\caption{The ratio $\rho/T^3 \propto \mu P/T^4$ is approximately constant
in both the 1.1 \Msol{} models with, and without, diffusion and it has
close to the same value in both.  See the text for a discussion of its
role in the appearance of convective cores.
}
\label{fig:pt4}
\end{figure}

\begin{figure}
\centerline{
\includegraphics[width=\textwidth]{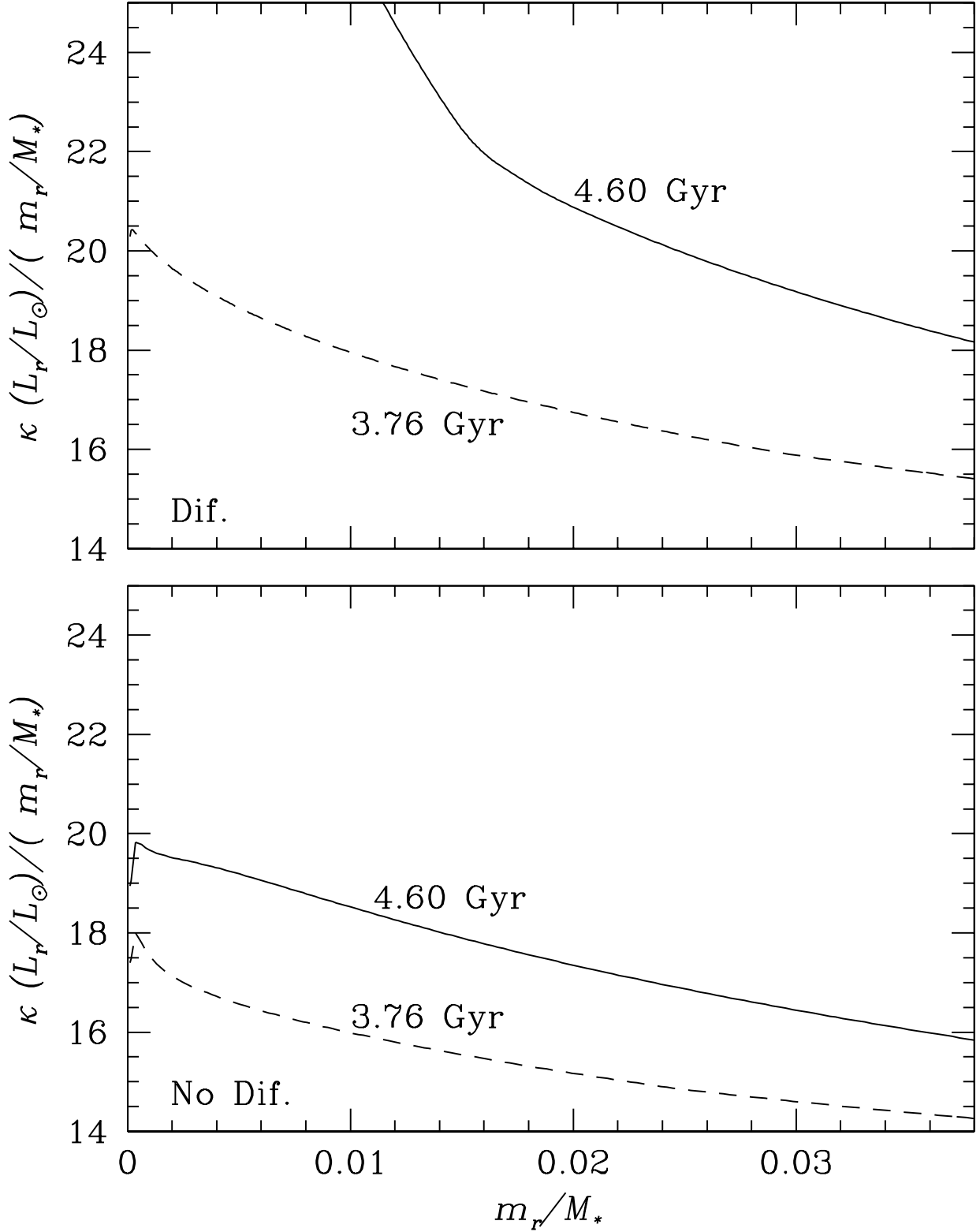}}
\caption{The variation of the ratio $\kappa L_r/m_r$ with mass and with
age near the centers of the 1.1 \Msol{} models with, and without, diffusion.
Note that this ratio is larger in the model with diffusion at a given age.
See the text for a discussion of its role in the appearance of a central
convective core.
}
\label{fig:klmt}
\end{figure}

The differences in chemical composition, and hence in opacity, appear to
be the main cause of the structural differences between the models with,
and without, diffusion.  The most evident difference is the convective
core that appears shortly after 3.76 Gyr in the diffusive models, but not
in the non-diffusive one.  In the bottom row of panels in
Figure \ref{fig:kcentre1} is plotted, as a function of the fractional mass,
$\nabla \rad - \nabla \ad$, where (from \citealt{Cox68}, Eq.~23.171):
\begin{equation}
\label{eq:thermal}
 \nabla \rad =  d\ln T/ d\ln P =
               \frac{3}{16\pi a c G} \frac{P}{T^4} \frac{\kappa L_r}{m_r}.
\end{equation}
According to Eq.~(2.5) and (2.8) of \citet{Stein66}, one may expect
the product $T^3/\rho$ to be approximately constant: this is seen
in Figure \ref{fig:pt4} to hold reasonably well in both the diffusive
and non-diffusive models.  The $P/T^4$ term then varies as $\mu^{-1}$
and so decreases as $Y$ increases.  (For instance, $Y$ increases over time from
0.53 to 0.61 at $m_r/\Mstar$ = 0.038 in the non-diffusive model.)  The
increasing $Y$ also causes a decrease in the opacity, as may be seen in
Fig.~\ref{fig:kcentre1}.  At a given $m_r$, the ratio
$L_r/m_r$ increases with time until hydrogen is exhausted.  Thus, there
are two partially cancelling effects in $\kappa L_r/m_r$ which, however, turns
out to increase with time (see Fig.~\ref{fig:klmt}).

At 3.76 Gyr, the expression $\kappa L_r/m_r$ is 10\%
larger in the diffusive, than in the non-diffusive, model, of which
4\% comes from the larger $\kappa$, and 6\% is due to the larger
value of $L_r/m_r$ in the diffusive model. 
At that phase, it is apparent that $\nabla \rad - \nabla \ad$ is close to
zero, especially in the case of the diffusive models.  For them,
$d\ln T/ d\ln P$ continues to increase and a convective core appears.
However, the central region of the non-diffusive model remains radiative.
The reason for this difference is, then, that the small metallicity-induced
opacity enhancement in the central region of the 1.1 \Msol{} model
with diffusion is large enough for a convective core to appear in this
model (but not in the one without diffusion) before the opacity is reduced
too much by the increasing He abundance.
 
In the preceding discussion, we implicitly used Schwarzschild's
stability criterion, although the calculations were done using
the Ledoux stability criterion.  The use of the latter rather than the
former has only a moderate effect on the size of the convective core.
The main effect of the Ledoux criterion is to temporarily transform a
convection zone into a semiconvection zone.  Semiconvection then mixes
He, however, thereby eliminating the $\mu$ gradient and the Schwarzschild
criterion is recovered over most of the convective core (as may be seen
in Figure \ref{fig:kcentre1}). There remains an extension of about 20\%
of the convective core caused by semiconvection.  However, the total
size of the convective core and of its semiconvective extension
approximately equals the size of the convective core that would be
obtained if the Schwarzschild criterion were used instead of the
Ledoux criterion (since $\nabla \rad = \nabla \ad$ approximately at the
outer boundary of the semiconvective core; see the lower part of
Fig.~\ref{fig:kcentre1}).  Note that allowing for the diffusion of the
metals leads to another 15--20\% increase in the size of the core
(compare the core mass in the model with the diffusion of He only
to that obtained when the diffusion of metals is also treated). 

\section{Isochrones}
\label{sec:iso}
\begin{figure}
\centerline{
\includegraphics[width=\textwidth]{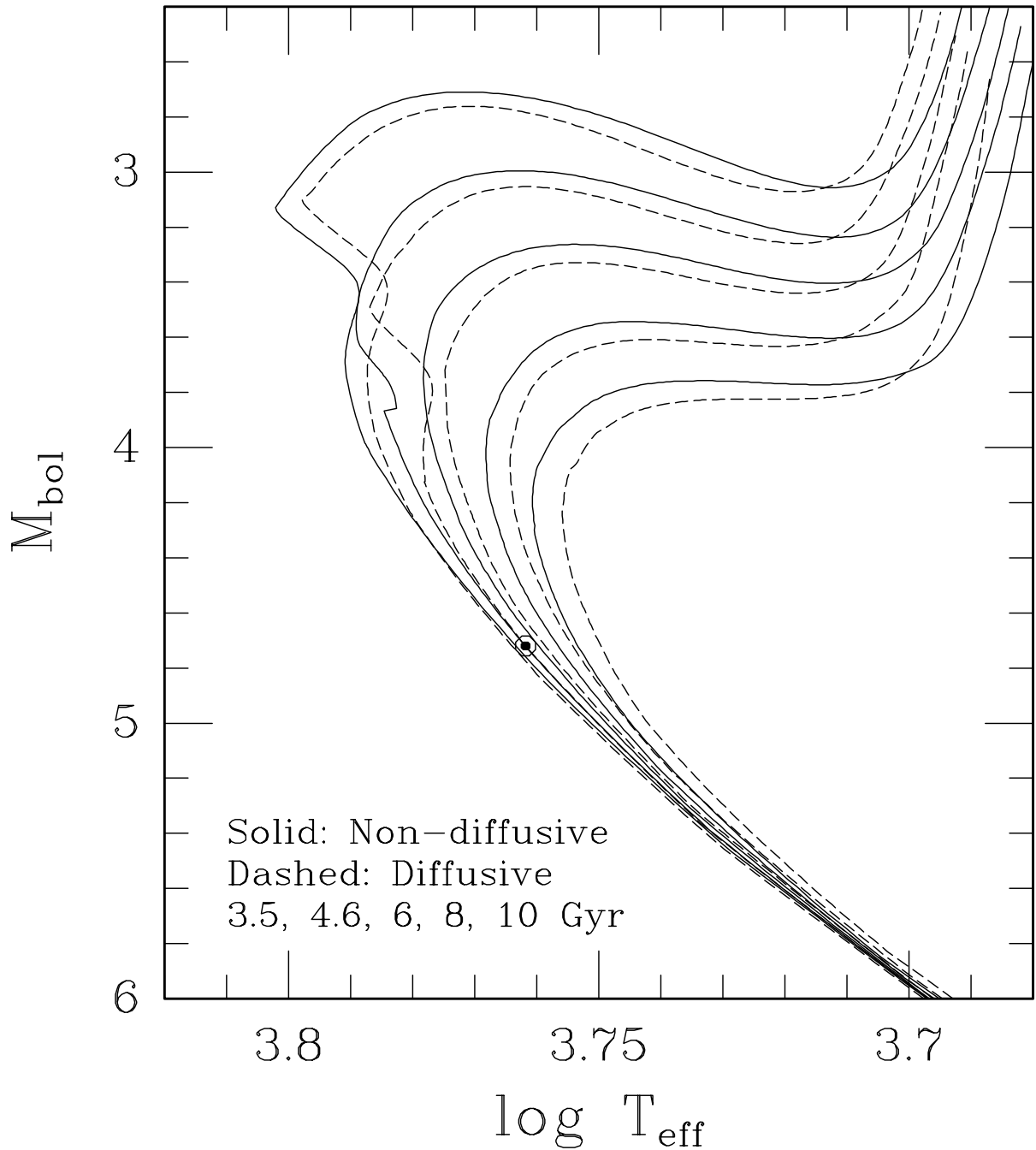}}
\caption{Comparison of non-diffusive and diffusive isochrones ({\it solid} and
{\it dashed curves}, respectively) for [m/H] $= 0.0$ and the indicated ages.
The location of the Sun on this diagram is given by the {\it solar symbol}.
To satisfy the solar constraint, the non-diffusive and diffusive isochrones
had to be shifted by $\delta\log T_{\rm eff} = -0.0022$ and $-0.0012$,
respectively).}
\label{fig:isochr1}
\end{figure}

The interpolation code described by \citet{BergbuschVa92} has been
used to generate isochrones for ages from 3.5 to 10 Gyr from both
the non-diffusive and diffusive grids of evolutionary tracks.
Figure~\ref{fig:isochr1} illustrates several of the computed isochrones
and shows that, at the same age, the diffusive isochrones have cooler
turnoffs, fainter subgiant branches, and bluer giant branches than those
which neglect gravitational settling and radiative accelerations --- even
when both sets of models are precisely normalized to the Sun, as indicated.
(In order to satisfy the solar constraint, the diffusive models required
a higher value of the mixing-length parameter, which is the main cause
of the differences between the {\it dashed} and {\it solid curves} at
the base of the red-giant branch.  See \S \ref{sec:calcul} and Table \ref{tab:parameters})

\begin{figure}
\plotone{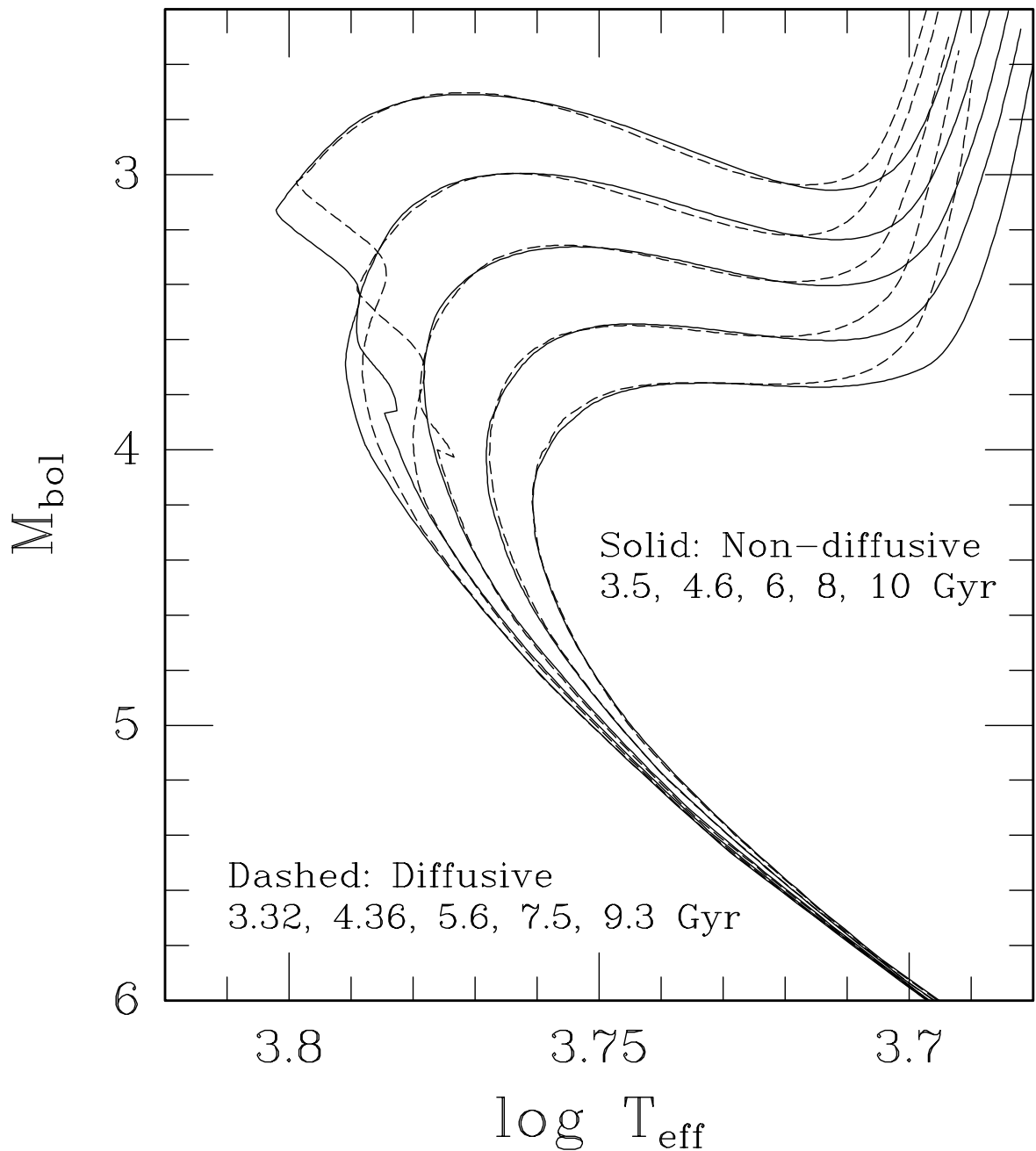}
\caption{As in the previous figure, except that diffusive isochrones have been
selected so as to provide the closest match to the subgiant branches and/or the
turnoffs of the non-diffusive isochrones.  The ages of the former are
$\sim 5-7$\% less than the latter.  Note that, in order to match the 
main-sequence locations of the non-diffusive isochrones, the diffusive
isochrones were shifted by small amounts ranging from
 $\delta \log T_{\rm eff}= -0.001$ to $+0.002$ over the age range considered.} 

\label{fig:isochr2}

\end{figure}

A more interesting and informative comparison of the isochrones is given in
Figure~\ref{fig:isochr2}.  In this case, isochrones from the non-diffusive and
diffusive grids are plotted that resemble each other most closely; i.e., they
predict very similar turnoff and subgiant luminosities.  Allowance for
diffusive processes clearly leads to a 5--7\% reduction in age at a given
turnoff luminosity (over the age range considered), which is considerably less
than the 10--12\% reduction that is predicted by models for extreme Population 
II stars (see \citealt{VandenBergRiMietal2002}).  However, the latter models
were constructed for the {\it same} initial helium and heavy-element
abundances, whereas the present computations have assumed {\it different}
initial values of $Y$ and $Z$ in order that both the non-diffusive and
diffusive models for $1.0 \Msol{}$ satisfy the solar contraint.  It is, in
fact, the differences in the assumed chemistry of the respective Standard
Solar Models that have compensated for nearly half of the expected effects
of diffusion on predicted turnoff luminosity--age relations.

Fig.~\ref{fig:isochr2} also shows that atomic diffusion has important
ramifications for the morphology of isochrones in the vicinity of the
turnoff.  In particular, the ``hook" feature in the youngest isochrones, 
which traces the rapid contraction phase that occurs at central H exhaustion
in those stars that have convective cores during their \MS{} phase, is
displaced to somewhat higher luminosities and cooler temperatures when
diffusive processes are treated.  (Note that the largest differences
between the {\it solid} and {\it dashed} loci occur when they deviate to
higher values of $\teff$ just prior to the beginning of the subgiant stage.)
This is very reminiscent of the effects of convective core overshooting.
(Indeed, as already mentioned, diffusive models {\it do} have enlarged convective cores and, as
shown in the next section, they provide a much improved match to the CMD
of the $\approx 4$ Gyr, open cluster M$\,$67, as compared with those
that neglect diffusion and convective overshooting.)

Moreover, convective cores clearly persist to fainter absolute magnitudes
when diffusion is treated: at the same turnoff luminosity, the diffusive
isochrone for 5.6 Gyr possesses a small ``hook" feature, while none is
present in the 6.0 Gyr non-diffusive isochrone.  In fact, the maximum age 
for which an observed CMD is expected to show a gap near the main-sequence
turnoff is $\approx 5.7$ Gyr if diffusion is treated, and $\approx 4.8$ Gyr
if diffusion is neglected.  Thus, open clusters with ages between approximately
4.8 and 5.7 Gyr have the potential to further test the effects of diffusion
physics in the central regions of stars.  (This difference in age is a
consequence of the fact that the stellar mass marking the transition
between tracks that have convective cores throughout the \MS{} phase, and
those which do not, is lower for the diffusive models --- $1.097 \Msol{}$
versus $1.14 \Msol{}$, see \S~\ref{sec:CZ}.)

\begin{figure}
\plotone{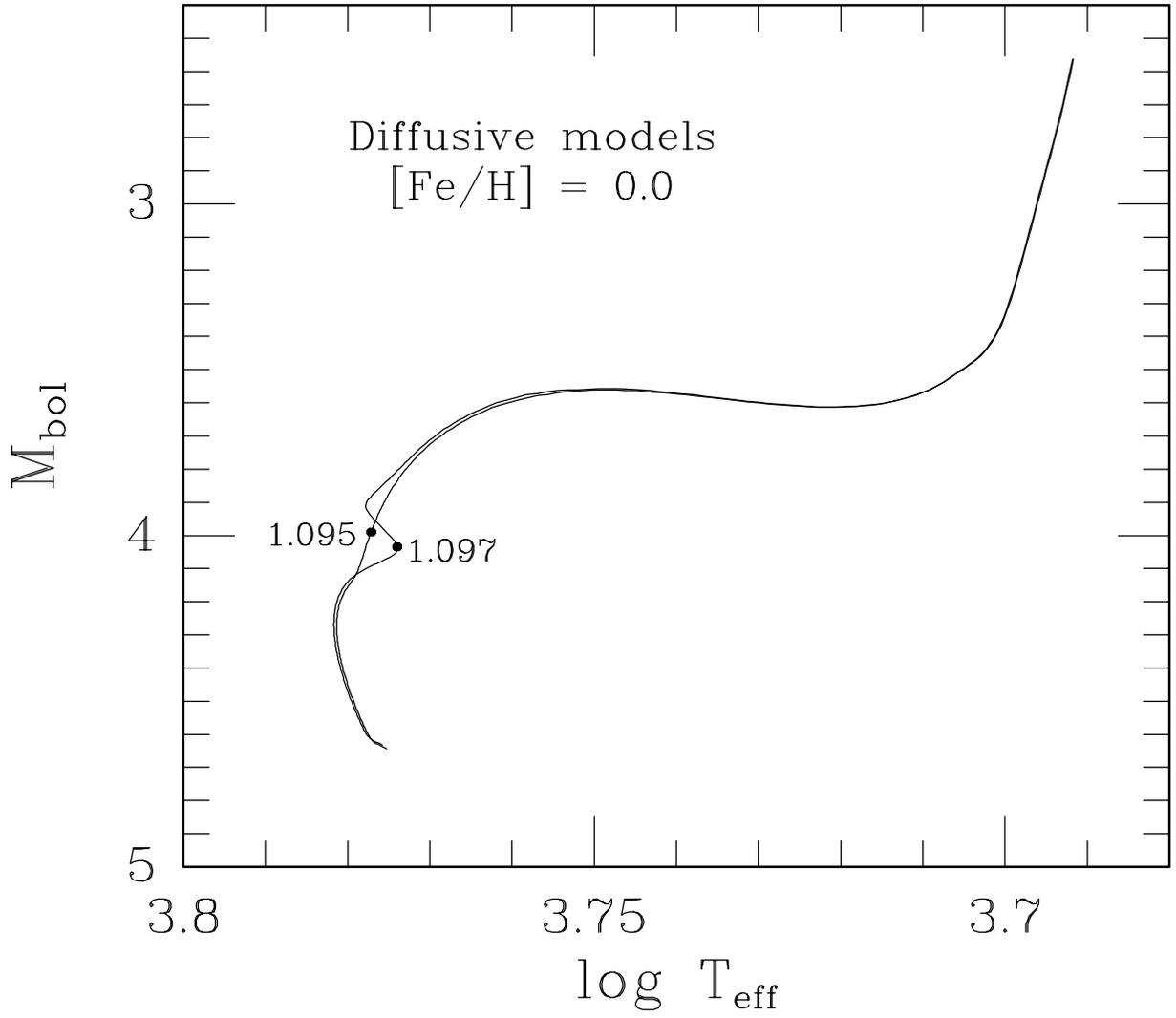}
\caption{Evolutionary tracks from the zero-age main sequence to the lower
giant branch for 1.095 and $1.097 \Msol{}$ and the solar metallicity, to
show the very rapid devlopment of the convective hook feature.  Diffusive
processes are treated in these computations.}
\label{fig:transmass}
\end{figure}

As shown in Figs.~\ref{fig:isochr1} and \ref{fig:isochr2}, the blueward
hooks in the oldest isochrones that possess such features (in both the
diffusive and non-diffusive grids) have small ``kinks" at their faint ends.
They arise because of the sudden change in the track morphology at the
transition mass.  Consider, for instance, the tracks plotted in
Figure~\ref{fig:transmass} for 1.095 and $1.097 \Msol{}$ stars, in the case
that diffusion is treated.  The {\it filled circles} indicate where the
predicted age is 5.55 Gyr on both tracks and it is clear that an isochrone
for this age (and similar ages) must undergo a redward jog between these
two points. 

\section{Application to M$\,$67 and NGC$\,$188}
\label{sec:opencl}
In their presentation of improved $UBVI$ photometry for M$\,$67,
\citet{MeynetMeMa93} concluded that ``none of the current isochrones
fit our data consistently''. The morphology of the main-sequence
turnoff and the luminosity of the gap at the turnoff were especially
problematic for the models that they considered.  Since that study
was published, the evidence has become overwhelming that there
is significant overshooting beyond the boundaries of convective cores as
determined from the Schwarzschild criterion (e.g., \citealt{MeynetMeMa93};
\citealt{DemarqueSaGu94}; \citealt{NordstroemAnAn97};
\citealt{SchroderPoEg97}; \citealt{RosvickVa98}).  There has also been
widespread agreement that the amount of overshooting is less in stars
that are just above the mass marking the transition between stars that
possess convective cores on the \MS{} and those which do not, than in stars
of appreciably higher mass. In particular, the extent of core overshooting
appears to be equivalent to $\approx 0.1$ pressure scale heights in the
turnoff stars of M$\,$67, whereas something closer to $0.25 H_P$ is typically
found in studies of much younger open clusters (see the aforementioned
papers, as well as \citealt{SarajediniHiKoetal99}).  However, given the
results described in the previous section, is it possible that diffusive
isochrones can provide a good fit to the M$\,$67 CMD without requiring
{\it any} convective overshooting?

To answer this question, we have transposed our isochrones to the
observed plane using the semi-empirical color--$\teff$ relations
described by \citet{VandenbergCl2003}, and performed a main-sequence fit
of the \citet{MontgomeryMaJa93}  CMD for M$\,$67 to the isochrones.
The assumption of a solar metallicity is within the $1\sigma$
uncertainty of most estimates of the cluster [m/H] value.  For instance,
Sarajedini et al.~(1999) concluded that M$\,$67 has [m/H] $=-0.05\pm
0.08$ from their consideration of the best available determinations prior to
their paper, and the latest high-resolution spectroscopic study that we are
aware of has obtained [m/H] $= -0.03\pm 0.03$
(\citealt{TautvaisienneEdTuetal2000}).  Moreover, the reddening that is
obtained from the \citet{SchlegelFiDa98} dust maps, $E(B-V) = 0.038$,
is in very good agreement with independent estimates (see the Sarajedini et
al.~study).  Consequently, the distance modulus that is derived from a 
main-sequence fit to the isochrones should be quite accurate (under these
assumptions).

\begin{figure}
\plotone{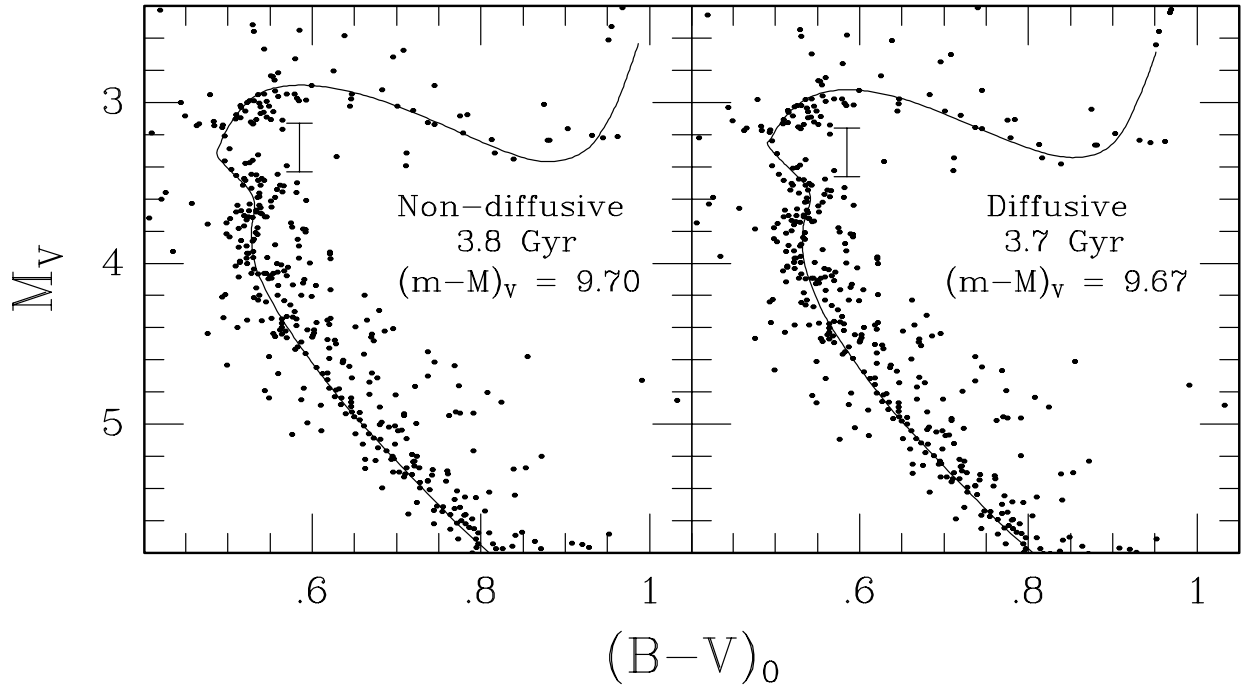}  
\caption{Main-sequence fits of the \citet{MontgomeryMaJa93}  CMD for M$\,$67 to
the non-diffusive and diffusive isochrones that provide the best match to the
cluster's subgiant branch.  The reddening is assumed to be $E(B-V) = 0.038$ mag
(Schlegel et al.~1998), and the derived distance modulus is $(m-M)_V = 9.70$
and 9.67, respectively.  Note the differences in the vicinity of the turnoff,
which indicate a clear preference for the diffusive isochrone.  Our estimate
of the luminosity spanned by the gap in M$\,$67 is indicated by the vertical
line bounded by short horizontal lines: it is used in the next plot.}
\label{fig:m67bvfit}
\end{figure}

The left-hand panel of Figure~\ref{fig:m67bvfit} shows how well the
non-diffusive isochrones are able to reproduce the M$\,$67 CMD.  The derived
distance modulus is $(m-M)_V = 9.70$ and the age of the isochrone that
provides the best match to the cluster subgiants is 3.8 Gyr.  M$\,$67 is known
to have a high binary fraction --- \citet{MontgomeryMaJa93}  have estimated
that at least 63\% of the cluster stars are binaries --- which certainly
complicates the interpretation of the data.  For instance, a large fraction
of the group of stars just above the gap (at $M_V\approx 3.1$) are likely to
be binaries given that such a large number of stars at nearly the same color
on the subgiant branch is contrary to the predictions of stellar evolutionary
theory.  The fact that they are displaced by 0.5--0.75 mag above the
main-sequence population is consistent with many of them being nearly
equal-mass binaries (see the simulated CMDs reported by
\citealt{CarraroChBretal94}).

In most respects, the isochrone fits the observed CMD rather well.  However,
the predicted location of the termination of the main-sequence, and hence of the
gap just above it, are somewhat too faint.  As illustrated in the right-hand
panel of Figure \ref{fig:m67bvfit}, this difficulty can be alleviated to
some extent if the observations are fitted to diffusive isochrones.  In this
case, a slightly smaller distance modulus, $(m-M)_V = 9.67$, is obtained from
the main-sequence fit, and the inferred age is also slightly less (3.7 Gyr).
(If the same distance modulus were adopted as in the left-hand panel, the
inferred age would be closer to 3.6 Gyr.)  Although the comparison between
theory and observation is still not completely satisfactory (the isochrone
appears to be a bit too red at $3.6\lta M_V\lta 4.1$), it does represent
a significant improvement over that given in the left-hand panel.  (Even the
predicted location of the base of the red-giant branch is much more consistent
with that observed.)

\begin{figure}
\plotone{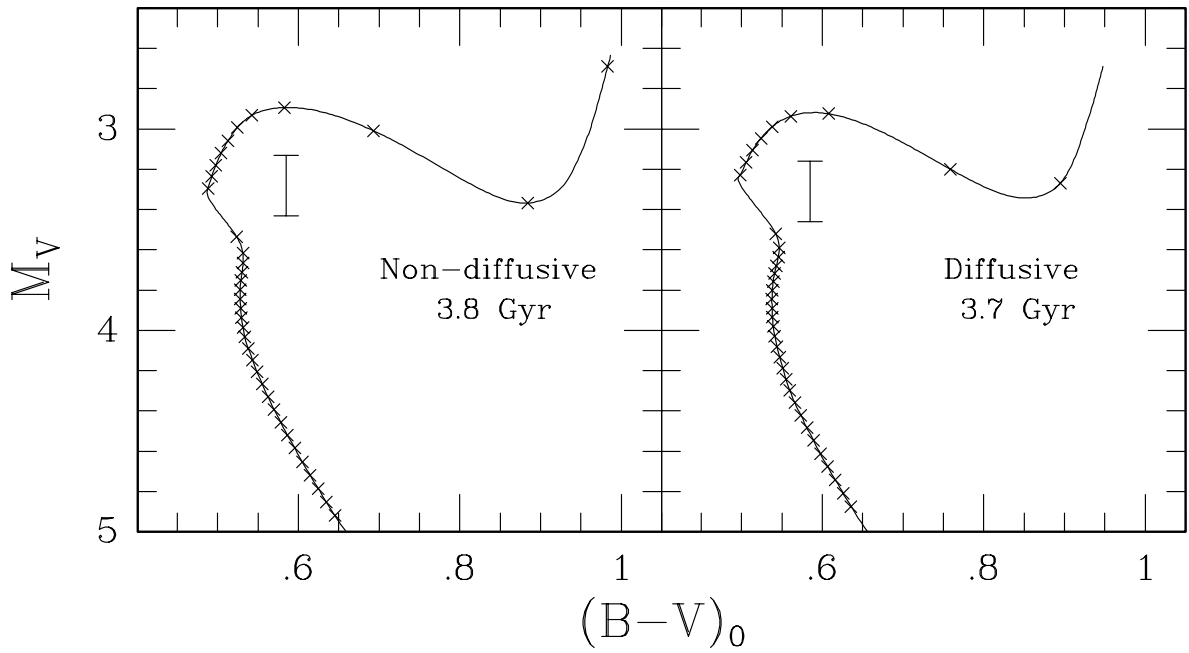}
\caption{The isochrones from the previous figure are plotted with crosses
superposed at $0.01 \Msol{}$ mass intervals.  To first approximation, the
same number of cluster stars is expected between adjacent crosses.  The
vertical line bounded by short horizontal lines indicates the magnitude range
spanned by the gap in M$\,$67.  It is to be compared with the location of
the predicted gap (as defined by the blue hook) in both panels.}
\label{fig:massiso}
\end{figure}

To reinforce this conclusion, we show in Figure~\ref{fig:massiso} the same
isochrones that appear in the previous figure with crosses plotted along them
at $0.01 \Msol{}$ intervals.  The density of the crosses gives a good
indication of the expected variation in the numbers of stars along the two
isochrones.  For instance, the blueward hook should manifest itself as a gap
in the distribution of turnoff stars, and relatively few stars should be
found on the subgiant branch because the rate of evolution is fast, and the
variation of mass with evolutionary state is low, in this phase.  The vertical
line bounded by short horizontal lines just to the right of each isochrone
indicates the observed location of the gap in M$\,$67 (from
Fig.~\ref{fig:m67bvfit}).  Given that considerably fewer stars are predicted
to be found in the magnitude range encompassed by the observed gap in the
right-hand panel than in the left-hand panel, the diffusive isochrone
clearly provides the best fit to the observations.  (Whether or not the
model fit could be further improved by assuming a small amount of convective
overshooting is difficult to say in view of the high fraction of binary
stars and significant field star contamination.)

It is, of course, very comforting that the diffusive models appear to be the
most realistic ones since it is well known that such calculations are favored
from solar oscillation studies --- e.g., \citet{ChristensenDalsgaardPrTh93};
\citet{RichardVaChetal96}.  At this time, we can only speculate that errors
in the adopted color--$\teff$ relations or the assumed abundances (perhaps of
helium) are responsible for the small color offset between the models and
observations at $M_V\sim 3.8$.

\begin{figure}
\plotone{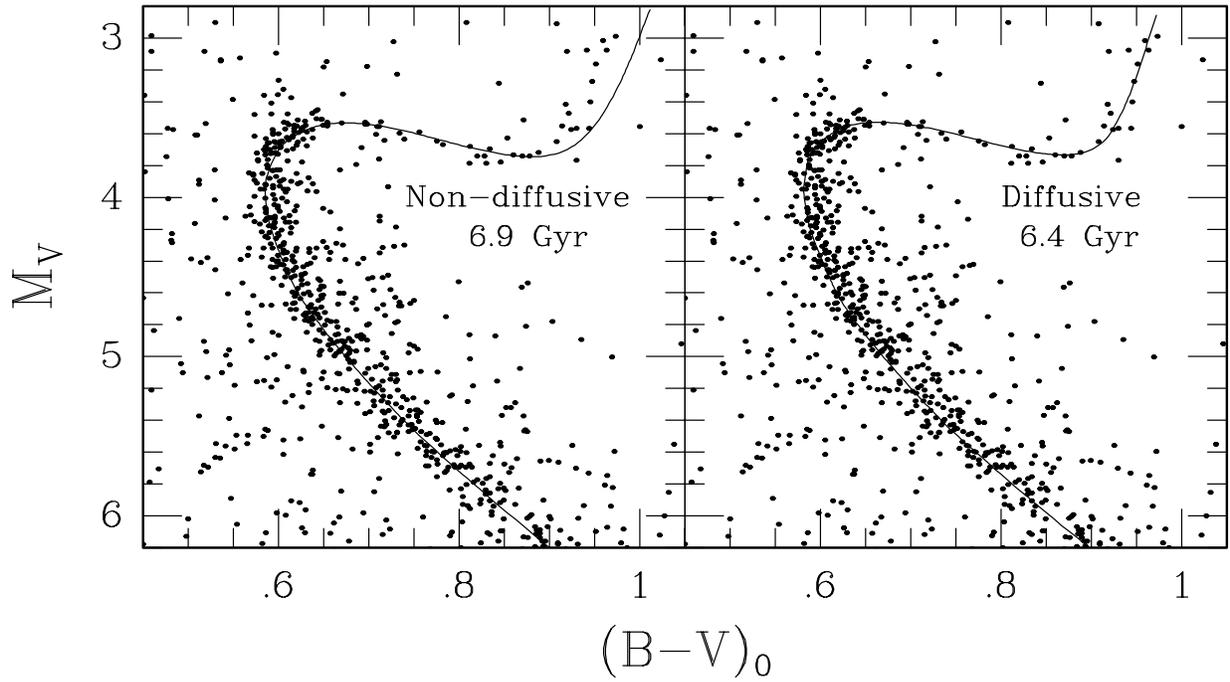}
\caption{Similar to Figure~\ref{fig:m67bvfit}, except that the CMD of NGC$\,$188
(Sarajedini et al.~1999) is fitted to non-diffusive and diffusive isochrones
that provide the best match to the cluster's subgiant branch, if it is assumed
that $E(B-V) = 0.087$ (Schlegel et al.~1998).  In both cases, the derived
distance modulus is $(m-M)_V = 11.40$.}
\label{fig:n188bvfit}
\end{figure}

As noted above, a solar abundance, open cluster having an age between 4.8
and 5.7 Gyr would provide a good test of the models since it is only the
diffusive models in this age range that predict the existence of a
main-sequence gap.  Unfortunately, only a few old open clusters have been
identified to date, and it seems unlikely that any of them have the right
age to provide such a test.  Perhaps the best candidate is NGC$\,$188, but
it appears to be too old by $\sim 0.5$--1 Gyr.  In the left-hand panel of
Figure~\ref{fig:n188bvfit}, a main-sequence fit of the Sarajedini et
al.~(1999) CMD for NGC$\,$188 to the non-diffusive isochrones yields
$(m-M)_V = 11.40$ and an age of 6.9 Gyr, on the assumption of $E(B-V)
= 0.087$ (Schlegel et al.~1998).  As noted by Sarajedini et al., this
reddening estimate is in good agreement with independent determinations,
and there is considerable spectroscopic support for a metallicity near
solar.  They adopted [m/H] $= -0.04\pm 0.05$, but more recent work
(\citealt{RandichSePa2003}; \citealt{WortheyJo2003}) favors [Fe/H] $\gta 0.0$.
The isochrone provides quite a satisfactory fit to the observations, except
at the base of the red-giant branch.

If the same reddening is assumed, a main-sequence fit of the photometry to
the diffusive isochrones also yields $(m-M)_V = 11.40$, but an age of 6.4 Gyr.
As shown in the right-hand panel of Figure \ref{fig:n188bvfit}, this isochrone
provides a very good match to the observed CMD, including the lower giant
branch.  Thus, by treating gravitational settling and radiative accelerations,
the inferred age of NGC$\,$188 has been reduced by $\approx 7$\%.  There is
no indication of a gap in the observed CMD, nor is any predicted, but it is
curious that the majority of the stars at $3.8\lta M_V \lta 4.2$ are redder
than the isochrone (in both panels), giving one the impression that the
best-fitting isochrone should have a small redward jog in this magnitude range.

\section{Conclusions}
\label{sec:conclusion}
Since the Sun is used to normalize convection parameters and initial
abundances, one could have imagined that, in solar metallicity clusters
having ages similar to that of the Sun, models with diffusion would lead
to the same age and the same CMD properties as models without diffusion.
The variations of $\alpha$ and initial abundances required to fit the Sun
in the diffusion model reproduce the same 1.0 \Msol{} star at the same age
and the two sets of models could be expected to do the same for star clusters.
Reality turns out to be more complex.  For age determinations, partial
cancellation effectively occurs, but the shapes of isochrones turn out
to be quite different near the turnoff.  Both the normalization to solar
abundances and the additional gravitational settling in the central regions
of stars work together to cause an 18\,\% increase in the central metallicity.
This increases the size of the convective core in stars of 1.09 to 1.3
\Msol{} (see \S \ref{sec:CZ}) which, in turn, modifies the morphologies of
isochrones (see \S \ref{sec:iso}) around the solar age.

An important consequence of the changes in the shapes of isochrones that
arise when diffusive processes are treated is that it is possible to match
the CMD of M$\,$67 (including the luminosity of the gap near the turnoff)
without having to assume an {\it ad hoc} amount of convective core
overshooting: a diffusive isochrone for 3.7 Gyr does a remarkably good job
of matching the cluster observations.  The other significant result of this
investigation, as far as isochrones are concerned, is that a gap near the
turnoff is predicted to persist in open clusters up to an age of $\approx
5.7$ Gyr by the diffusive models, whereas the limiting age is closer to 4.8
Gyr if diffusion is not treated.  It would be important to have detailed
observations of the fiducial sequences for such clusters as those identified
by \citet{FrielJaTaetal2002} to test this prediction.  Unfortunately,
NGC$\,$188 appears to be too old to do this, given that our best estimate
of its age is 6.4 Gyr based on the diffusive isochrones (which, incidently,
provide a superb match to the observed CMD).

The predicted surface abundance variations among near turnoff stars turn out
to be limited to approximately 0.1 dex in M\,67 and 0.07 dex in NGC\,188 (see \S \ref{sec:composition}). Most
elements heavier than Si have their surface abundances modified by \gr{}
but no large overabundances are expected.  While not
negligible, such variations are not easy to detect at the present time.  The
existing Li/Be measurements in a few stars of M\,67  (see \S \ref{sec:li_be}) suggest 
that another process may be required to reduce the Li abundance (see \citealt{SillsDe2000})
though the Be/Li trend obtained with models including all aspects of atomic diffusion is very different 
from the trend obtained by these authors.  
  With improved observations this  becomes a test of various turbulent models and will be further discussed
in a paper in preparation on LiBeB abundances in cluster and field \MS{} stars.

\acknowledgements
This work has been supported by  Operating Grants to G.~M.~and to
D.~A.~V.~from the Natural Sciences and Engineering Research Council of Canada.  
We thank the R\'eseau Qu\'eb\'ecois de Calcul de Haute Performance (RQCHP)
for providing us with the computational resources required for this work.

\newpage

\newpage

\clearpage

\clearpage

\clearpage

\clearpage

\clearpage
\begin{center}
\begin{deluxetable}{llllllll}
\footnotesize
\tablecaption{Computed Solar Metallicity Tracks}
\tablecolumns{8}
\tablehead{ \colhead{$Z_{\rm 0}$} &
\colhead{$Y_{\rm 0}$} &
\colhead{$\alpha_{\rm MLT}$} & \colhead{Boundary\tablenotemark{a}} & 
\colhead{Atomic} &
\colhead{Turbulence} & \colhead{$L/L_{\odot}$\tablenotemark{b}} &
\colhead{$R/R_{\odot}$\tablenotemark{c}} \\
 & & & condition & diffusion & & & \\}
\startdata
$0.01750$  & $0.26811$  & $1.94646$ & KS & No  & No          & $0.999$ & 
$0.996$ \\
$0.01999$  & $0.27769$  & $2.09635$ & KS & yes & No          & $1.002$ & 
$1.002$ \\
$0.01999$  & $0.27769$  & $2.09635$ & KS & yes & T6.09\tablenotemark{d} & $1.001$ & 
$1.002$ \\
\enddata
\tablenotetext{a}{\;KS: Krishna-Swamy}
\tablenotetext{b}{\;$L_{\odot}=3.86 \times 10^{33}$ erg.s$^{-1}$}
\tablenotetext{c}{\;$R_{\odot}=6.9599 \times 10^{10}$ cm}
\tablenotetext{d}{\;See \citet{RichardMiRietal2002}}
\label{tab:parameters}
\end{deluxetable}
\end{center}

\clearpage

\end{document}